\documentclass[onecolumn,11pt]{revtex4}

\usepackage{graphicx}
\usepackage{dcolumn}
\usepackage{bm}
\usepackage{color}
\usepackage{amssymb,amsmath}

\input epsf

\begin{document}

\title{\Large Constructions of $f(R,G,\mathcal{T})$ Gravity from Some Expansions of the Universe}

\author{\bf Ujjal Debnath\footnote{ujjaldebnath@gmail.com} }

\affiliation{Department of Mathematics, Indian Institute of
Engineering Science and Technology, Shibpur, Howrah-711 103,
India.}

\begin{abstract}
Here we propose the extended modified gravity theory named as
$f(R,G,\mathcal{T})$ gravity where $R$ is the Ricci scalar, $G$ is
the Gauss-Bonnet invariant and $\mathcal{T}$ is the trace of the
stress-energy tensor. We derive the gravitational field equations
in $f(R,G,\mathcal{T})$ gravity by taking least action principle.
Next we construct the $f(R,G,\mathcal{T})$ in terms of $R$, $G$
and $\mathcal{T}$ in de Sitter as well as power law expansion. We
also construct $f(R,G,\mathcal{T})$ if the expansion follows the
finite time future singulary (big rip singularity). We investigate
the energy conditions in this modified theory of gravity and
examine the validities of all energy conditions. Finally,
we analyze the stability of the constructed modified gravity. \\

\noindent {Keywords:} $f(R,G,\mathcal{T})$ gravity, energy
conditions, stability.
\end{abstract}

\maketitle

\section{Introduction}

Recent observational data suggests that our universe is
experiencing an accelerated expansion
\cite{Riess,Perl,Sper1,Sper2}. This acceleration is caused by some
unknown matter known as dark energy which has the property that
positive energy density ($\rho$) and negative pressure ($p$)
satisfying $\rho+3p<0$. It is believed that our present Universe
is made up of about 4\% ordinary matter, about 74\% dark energy
and about 22\% dark matter. A simple candidate for dark energy is
the cosmological constant \cite{Pad,Sahni}. Also several models
play the roles of dark energy such as quintessence \cite{Ratra},
phantom \cite{Cald1}, quintom \cite{Feng,Guo}, tachyon \cite{Sen},
k-essence \cite{Arme}, dilaton \cite{gas}, hessence \cite{Wei},
DBI-essence \cite{Gum,Mart} etc. There are other alternative of
dark energy is modified gravity theory \cite{Sah,Noj} which
represents a classical generalization of general relativity. In
theoretical model, the standard Einstein-Hilbert action is
replaced by different functions of the Ricci scalar $R$
\cite{Bam,Cap} (i.e. $f(R)$ gravity) or Gauss-Bonnet invariant $G$
\cite{Sam,Noj1} (i.e., $f(G)$ gravity). These modifications should
consistently describe the early-time inflation and late-time
acceleration, without introduce any other dark component and
consistent with the solar system constraints \cite{Felic}. A
generalization of $f(R)$ modified gravity theory was proposed by
Bertolami et al \cite{Ber} by including an explicit coupling of
arbitrary function of the Ricci scalar $R$ with the matter
Lagrangian $L_{m}$. Nojiri et al \cite{Noj2} studied the
non-minimally coupling of $f(R)$ and $f(G)$ gravity theories with
$L_{m}$ and found that this coupling unifies the inflationary era
with recent cosmic accelerated expansion. The geodesic deviation
of $f(G)$ gravity (for small curvature) is weaker than the non-minimal $f(R)$ gravity.\\

Harko et al \cite{Harko} proposed another extensions of standard
general relativity, the $f(R,\mathcal{T})$ and
$f(R,\mathcal{T}^{\phi})$ modified theories of gravity, where the
gravitational Lagrangian is given by an arbitrary function of the
Ricci scalar $R$ and of the trace of the stress-energy tensor
$\mathcal{T}$ and $\mathcal{T}^{\phi}$ is the stress-energy tensor
of scalar field. The implications in $f(R,\mathcal{T})$ gravity
have been extensively studied in several works
\cite{Jam,Hou,Sh,Alv,My,Ch}. Another extension of modified gravity
is $f(R,G)$ gravity \cite{Bamba}. In $f(R,G)$ gravity, the energy
conditions, future finite time singularities and other
cosmological implications have been extensively studied by several
authors \cite{Fel,Fel1,Fel2,Fel3,Dom,Mak,Ata}. Recently, Sharif et
al \cite{Sharif} have introduced another kind of extension of
modified gravity theory like $f(G,\mathcal{T})$ gravity theory.
They have reconstructed the $f(G,\mathcal{T})$ gravity through
power law, de Sitter expansions of the Universe and also
investigated the validities of all the energy conditions. Also
Sharif et al \cite{Sharif1} have analyzed the stability of some
reconstructed cosmological models in $f(G,\mathcal{T})$ gravity.
Shamir et al \cite{Shamir} have studied the Noether symmetry
approach of some cosmologically viable $f(G,\mathcal{T})$ gravity
models. Motivated by these works, here we propose another
extension of modified theories of gravity like
$f(R,G,\mathcal{T})$ gravity and our main aim is to find the forms
of the function $f(R,G,\mathcal{T})$ from de Sitter, power law and
future singularity models. In section 2, we derive the
gravitational field equations in $f(R,G,\mathcal{T})$ gravity.
Next we construct the $f(R,G,\mathcal{T})$ in terms of $R$, $G$
and $\mathcal{T}$ in de Sitter expansion in section 3. We
construct the $f(R,G,\mathcal{T})$ due to the power law expansion
model in section 4. We also construct $f(R,G,\mathcal{T})$ in
section 5 if the expansion follows the finite time future
singulary. In section 6, we study the energy conditions in
modified theory of gravity. In section 7, we analyze the stability
of $f(R,G,\mathcal{T})$ gravity model. Finally we draw some
concluding remarks in section 8.

\section{Gravitational Field Equations in $f(R,G,\mathcal{T})$ Gravity}

Here we formulate the Einstein's field equations for
$f(R,G,\mathcal{T})$ modified gravity theory. For this purpose, we
consider the action for $f(R,G,\mathcal{T})$ gravity theory in the
form
\begin{equation}\label{1}
S=\frac{1}{2\kappa^{2}}~\int f(R,G,\mathcal{T})\sqrt{-g}~d^{4}x
+\int L_{m} \sqrt{-g}~d^{4}x
\end{equation}
where $f(R,G,\mathcal{T})$ is the arbitrary function of Ricci
scalar $R$, Gauss-Bonnet invariant $G$ and of the trace
$\mathcal{T}$ of stress-energy tensor of the matter. Also $L_{m}$
is the matter Lagrangian, $g=|g_{\mu\nu}|$ and $\kappa^{2}=8\pi G$
(choosing $c=1$). The stress-energy tensor of the matter is
defined as \cite{Land}
\begin{equation}\label{2}
T_{\mu\nu}=-\frac{2}{\sqrt{-g}}~\frac{\delta(\sqrt{-g}~L_{m})}{\delta
g^{\mu\nu}}
\end{equation}
The trace of the stress-energy tensor is
$\mathcal{T}=g^{\mu\nu}T_{\mu\nu}$. Here we assume that the matter
Lagrangian $L_{m}$ depends only on the metric tensor $g_{\mu\nu}$,
so we obtain
\begin{equation}\label{3}
T_{\mu\nu}=g_{\mu\nu}L_{m}-2~\frac{\partial L_{m}}{\partial
g^{\mu\nu}}
\end{equation}
Now the variation of action (\ref{1}), we obtain the following
integral:
\begin{equation}\label{4}
\delta S=\frac{1}{2\kappa^{2}}~\int \left[f_{R}\delta
R+f_{G}\delta G+f_{\mathcal{T}}\delta \mathcal{T}
+f\delta(\sqrt{-g})+\frac{2\kappa^{2}}{\sqrt{-g}}~\delta
(L_{m}\sqrt{-g}) \right]\sqrt{-g}d^{4}x
\end{equation}
where $f_{R}=\frac{\partial f}{\partial R}$, $f_{G}=\frac{\partial
f}{\partial G}$ and $f_{\mathcal{T}}=\frac{\partial f}{\partial
\mathcal{T}}$. The Ricci scalar $R$ and Gauss-Bonnet invariant $G$
are as follows:
\begin{equation}\label{5}
R=g^{\mu\nu}R_{\mu\nu},~G=R^{2}-4R_{\mu\nu}R^{\mu\nu}+R_{\mu\nu\xi\eta}R^{\mu\nu\xi\eta}
\end{equation}
The variations of $\sqrt{-g}$, $R$, $R_{\mu\nu}$, $G$,
$\mathcal{T}$ are as follows:
\begin{equation} \label{6}
\delta(\sqrt{-g})=-\frac{1}{2}~\sqrt{-g}~g_{\mu\nu}\delta
g^{\mu\nu},
\end{equation}
\begin{equation}
\delta
R=(R_{\mu\nu}+g_{\mu\nu}\nabla^{2}-\nabla_{\mu}\nabla_{\nu})\delta
g^{\mu\nu},
\end{equation}
\begin{equation}
\delta
R_{\mu\nu}=\nabla_{\lambda}\delta\Gamma^{\lambda}_{\mu\nu}-\nabla_{\nu}\delta\Gamma^{\lambda}_{\mu\lambda},
\end{equation}
\begin{equation}
\delta\Gamma^{\lambda}_{\mu\nu}=\frac{1}{2}~g^{\lambda\alpha}(\nabla_{\mu}\delta
g_{\nu\alpha}+\nabla_{\nu}\delta
g_{\mu\alpha}-\nabla_{\alpha}\delta g_{\mu\nu}),
\end{equation}
\begin{equation}
\delta G=2R\delta
R-4\delta(R_{\mu\nu}R^{\mu\nu})+\delta(R_{\mu\nu\xi\eta}R^{\mu\nu\xi\eta}),
\end{equation}
\begin{equation}
\delta \mathcal{T}=(T_{\mu\nu}+\Theta_{\mu\nu})~\delta g^{\mu\nu}
\end{equation}
with
\begin{equation}\label{12}
\Theta_{\mu\nu}=g^{\alpha\beta}~\frac{\partial
T_{\alpha\beta}}{\partial g^{\mu\nu}}
\end{equation}
Now putting $\delta S=0$ in equation (\ref{4}) and using (\ref{6})
- (\ref{12}), we obtain the field equations of
$f(R,G,\mathcal{T})$ gravity as
$$
(R_{\mu\nu}+g_{\mu\nu}\nabla^{2}-\nabla_{\mu}\nabla_{\nu})f_{R}-\frac{1}{2}~fg_{\mu\nu}
+(2RR_{\mu\nu}-4R^{\xi}_{\mu}R_{\xi\nu}-4R_{\mu\xi\nu\eta}R^{\xi\eta}
+2R_{\mu}^{\xi\eta\lambda}R_{\nu\xi\eta\lambda})f_{G}
$$
$$
+(2Rg_{\mu\nu}\nabla^{2}-2R\nabla_{\mu}\nabla_{\nu}-4g_{\mu\nu}R^{\xi\eta}\nabla_{\xi}\nabla_{\eta}
-4R_{\mu\nu}\nabla^{2}+4R_{\mu}^{\xi}\nabla_{\nu}\nabla_{\xi}+4R_{\nu}^{\xi}\nabla_{\mu}\nabla_{\xi}
+4R_{\mu\xi\nu\eta}\nabla^{\xi}\nabla^{\eta})f_{G}
$$
\begin{equation}\label{13}
=\kappa^{2}T_{\mu\nu}-(T_{\mu\nu}+\Theta_{\mu\nu})f_{\mathcal{T}}
\end{equation}
where $\nabla^{2}=\nabla_{\mu}\nabla^{\mu}$ is the d'Alembert
operator. If we put $f(R,G,\mathcal{T})=f(R,\mathcal{T})$ ($G$
independent), we can recover the field equations in
$f(R,\mathcal{T})$ gravity which was proposed in Ref \cite{Harko}.
If we put $f(R,G,\mathcal{T})=f(G,\mathcal{T})$ ($R$ independent),
we can recover the field equations in $f(G,\mathcal{T})$ gravity
which was proposed in Ref \cite{Sharif} and if we put
$f(R,G,\mathcal{T})=f(R,G)$ ($\mathcal{T}$ independent), we can
recover the field equations in $f(R,G)$ gravity \cite{Bamba}.
Taking the trace of the above field equation (\ref{13})
(multiplying both sides by $g^{\mu\nu}$) we have
\begin{equation}
(R+3\nabla^{2})f_{R}-2f-(2G-2R\nabla^{2}+4R^{\mu\nu}\nabla_{\mu}\nabla_{\nu})f_{G}=\kappa^{2}\mathcal{T}-(\mathcal{T}+\Theta)f_{\mathcal{T}}
\end{equation}
where $\Theta=\Theta_{\mu\nu}g^{\mu\nu}$. Taking covariant
divergence of equation (\ref{13}), we obtain \cite{Harko,Sharif}
\begin{equation}\label{15}
\nabla^{\mu}T_{\mu\nu}=\frac{f_{\mathcal{T}}}{\kappa^{2}-f_{\mathcal{T}}}~\left[(T_{\mu\nu}+\Theta_{\mu\nu})\nabla^{\mu}~\ln
f_{\mathcal{T}}+
\nabla^{\mu}\Theta_{\mu\nu}-\frac{1}{2}~g_{\mu\nu}\nabla^{\mu}\mathcal{T}
\right]
\end{equation}
We see that the above expression is independent of $f_{R}$ and
$f_{G}$. Also we may obtain \cite{Harko,Sharif}
\begin{equation} \label{16}
\Theta_{\mu\nu}=-2T_{\mu\nu}+g_{\mu\nu}L_{m}-2g^{\alpha\beta}~\frac{\partial^{2}L_{m}}{\partial
g^{\mu\nu}\partial g^{\alpha\beta}}
\end{equation}
If $L_{m}$ is known then we can find $\Theta_{\mu\nu}$. The energy
momentum tensor for perfect fluid is assumed as
\begin{equation}
T_{\mu\nu}=(\rho+p)u_{\mu}u_{\nu}+pg_{\mu\nu}
\end{equation}
where $\rho$ and $p$ are respectively energy density and pressure
of perfect fluid. The four velocity $u_{\mu}$ satisfies
$u_{\mu}u^{\mu}=-1$ and $u^{\mu}\nabla _{\nu}u_{\mu}=0$. Now
assume that the matter Lagrangian is $L_{m}=p$. So the equation
(\ref{16}) reduces to
\begin{equation}
\Theta_{\mu\nu}=-2T_{\mu\nu}+pg_{\mu\nu}
\end{equation}
Here we assume the flat FRW model of the universe described by the
line element
\begin{equation}
ds^{2}=-dt^{2}+a^{2}(t)\left[dr^{2}+r^{2}(d\theta^{2}+sin^{2}\theta
d\phi^{2})\right]
\end{equation}
where $a(t)$ is the scale factor. From equation (\ref{5}), we can
obtain
\begin{equation}
R=6(\dot{H}+2H^{2}),~~G=24H^{2}(\dot{H}+H^{2})
\end{equation}
where $H=\dot{a}/a$ is the Hubble parameter and $dot$ means the
derivative w.r.t. cosmic time $t$. For the above metric, we obtain
the trace $\mathcal{T}=3p-\rho$ and $\Theta=2(\rho-p)$. So
$\mathcal{T}+\Theta=2(\rho+p)$. From equation (\ref{15}), we
obtain the non-conservation equation
\begin{equation}\label{21}
\dot{\rho}+3H(\rho+p)=\left(\frac{1}{2}\dot{\mathcal{T}}-\dot{p}\right)f_{\mathcal{T}}-(\rho+p)\dot{f}_{\mathcal{T}}
\end{equation}
Now, the standard conservation equation $\nabla^{\mu}T_{\mu\nu}=0$
for perfect fluid gives
\begin{equation}\label{22}
\dot{\rho}+3H(\rho+p)=0
\end{equation}
So from equation (\ref{21}), we obtain
\begin{equation}
(\dot{\rho}-\dot{p})f_{\mathcal{T}}+2(\rho+p)\dot{f}_{\mathcal{T}}=0
\end{equation}
From equation (\ref{13}) we obtain the field equations for
$f(R,G,\mathcal{T})$ gravity as
\begin{equation}\label{24}
3H^{2}=\frac{1}{f_{R}} \left[
\kappa^{2}\rho+(\rho+p)f_{\mathcal{T}}+\frac{1}{2}(Rf_{R}-f)-3H\dot{f}_{R}+12H^{2}(\dot{H}+H^{2})f_{G}-12H^{3}\dot{f}_{G}
\right]
\end{equation}
and
\begin{equation}\label{25}
(2\dot{H}+3H^{2})=-\frac{1}{f_{R}} \left[
\kappa^{2}p-\frac{1}{2}(Rf_{R}-f)+2H\dot{f}_{R}+\ddot{f}_{R}-12H^{2}(\dot{H}+H^{2})f_{G}
+8H(\dot{H}+H^{2})\dot{f}_{G}+4H^{2}\ddot{f}_{G} \right]
\end{equation}
The above two field equations can be written in the standard
Einstein's field equations as
\begin{equation}
3H^{2}=\kappa^{2}\rho_{eff}~~and~~
(2\dot{H}+3H^{2})=-\kappa^{2}p_{eff}
\end{equation}
where
\begin{equation}
\rho_{eff}=\frac{1}{\kappa^{2}f_{R}} \left[
\kappa^{2}\rho+(\rho+p)f_{\mathcal{T}}+\frac{1}{2}(Rf_{R}-f)-3H\dot{f}_{R}+12H^{2}(\dot{H}+H^{2})f_{G}-12H^{3}\dot{f}_{G}
\right]
\end{equation}
and
\begin{equation}
p_{eff}=\frac{1}{\kappa^{2}f_{R}} \left[
\kappa^{2}p-\frac{1}{2}(Rf_{R}-f)+2H\dot{f}_{R}+\ddot{f}_{R}-12H^{2}(\dot{H}+H^{2})f_{G}
+8H(\dot{H}+H^{2})\dot{f}_{G}+4H^{2}\ddot{f}_{G} \right]
\end{equation}
Now the first and 2nd derivatives of $f_{R}$ and $f_{G}$ w.r.t.
$t$ are given below:
\begin{equation}\label{29}
\dot{f}_{R}=\dot{R}f_{RR}+\dot{G}f_{RG}+\dot{\mathcal{T}}f_{R\mathcal{T}},~\dot{f}_{G}=\dot{R}f_{RG}+\dot{G}f_{GG}+\dot{\mathcal{T}}f_{G\mathcal{T}},
\end{equation}
\begin{equation}\label{30}
\ddot{f}_{R}=\ddot{R}f_{RR}+\ddot{G}f_{RG}+\ddot{\mathcal{T}}f_{R\mathcal{T}}+
\dot{R}^{2}f_{RRR}+2\dot{R}\dot{G}f_{RRG}+2\dot{R}\dot{\mathcal{T}}f_{RR\mathcal{T}}
+\dot{G}^{2}f_{RGG}+2\dot{G}\dot{\mathcal{T}}f_{RG\mathcal{T}}+\dot{\mathcal{T}}^{2}f_{R\mathcal{T}\mathcal{T}},
\end{equation}
\begin{equation}\label{31}
\ddot{f}_{G}=\ddot{R}f_{RG}+\ddot{G}f_{GG}+\ddot{\mathcal{T}}f_{G\mathcal{T}}+
\dot{R}^{2}f_{RRG}+2\dot{R}\dot{G}f_{RGG}+2\dot{R}\dot{\mathcal{T}}f_{RG\mathcal{T}}
+\dot{G}^{2}f_{GGG}+2\dot{G}\dot{\mathcal{T}}f_{GG\mathcal{T}}+\dot{\mathcal{T}}^{2}f_{G\mathcal{T}\mathcal{T}}
\end{equation}
Using (\ref{29})-(\ref{31}), the field equations (\ref{24}) and
(\ref{25}) reduce to the following equations
$$
\kappa^{2}\rho-\frac{1}{2}f+\frac{1}{2}(R-6H^{2})f_{R}+12H^{2}(\dot{H}+H^{2})f_{G}+(\rho+p)f_{\mathcal{T}}-3H\dot{R}f_{RR}
$$
\begin{equation}\label{32}
-3H(4H^{2}\dot{R}+\dot{G})f_{RG}-3H\dot{\mathcal{T}}f_{R\mathcal{T}}-12H^{3}\dot{G}f_{GG}-12H^{3}\dot{\mathcal{T}}f_{G\mathcal{T}}=0,
\end{equation}
and
$$
\kappa^{2}p+\frac{1}{2}f+\frac{1}{2}(4\dot{H}+6H^{2}-R)f_{R}-12H^{2}(\dot{H}+H^{2})f_{G}+(2H\dot{R}+\ddot{R})f_{RR}
$$
$$
+[8H(\dot{H}+H^{2})\dot{R}+4H^{2}\ddot{R}+2H\dot{G}+\ddot{G}]f_{RG}+(2H\dot{\mathcal{T}}+\ddot{\mathcal{T}})f_{R\mathcal{T}}
+[8H(\dot{H}+H^{2})\dot{G}+4H^{2}\ddot{G}]f_{GG}
$$
$$
[8H(\dot{H}+H^{2})\dot{\mathcal{T}}+4H^{2}\ddot{\mathcal{T}}]f_{G\mathcal{T}}+\dot{R}^{2}f_{RRR}+(2\dot{R}\dot{G}+4H^{2}\dot{R}^{2})f_{RRG}
+2\dot{R}\dot{\mathcal{T}}f_{RR\mathcal{T}}+(\dot{G}^{2}+8H^{2}\dot{R}\dot{G})f_{RGG}
$$
\begin{equation}\label{33}
+(2\dot{G}\dot{\mathcal{T}}+8H^{2}\dot{R}\dot{\mathcal{T}})f_{RG\mathcal{T}}+\dot{\mathcal{T}}^{2}f_{R\mathcal{T}\mathcal{T}}
+4H^{2}\dot{G}^{2}f_{GGG}+8H^{2}\dot{G}\dot{\mathcal{T}}f_{GG\mathcal{T}}+4H^{2}\dot{\mathcal{T}}^{2}f_{G\mathcal{T}\mathcal{T}}=0
\end{equation}

In the next sections, we'll construct the $f(R,G,\mathcal{T})$ in
terms of $R,~G$ and $\mathcal{T}$ for de Sitter, power law and
future singularity expansion models.

\section{Construction of $f(R,G,\mathcal{T})$ in de Sitter Model}

Now we want to construct the function $f(R,G,\mathcal{T})$ in
terms of $R,~G,~\mathcal{T}$ for de-Sitter universe. For de Sitter
model of the universe, the scale factor can be written as
\cite{Sharif}
\begin{equation}
a(t)=a_{0}e^{H_{0}t}
\end{equation}
where $a_{0}$ and $H_{0}$ are positive constants. For this model,
we have
\begin{equation}
H=H_{0},~R=12H_{0}^{2},~G=24H_{0}^{4},~\dot{R}=\dot{G}=\ddot{R}=\ddot{G}=0,
\end{equation}
Also we assume the fluid of the Universe obeys barotropic equation
of state $p=w\rho$, where $w$ is constant. So from the
conservation equation (\ref{22}), we obtain the energy density
$\rho=\rho_{0}a^{-3(1+w)}$, where $\rho_{0}$ is positive constant.
Now we obtain
\begin{equation}
\rho=\frac{\mathcal{T}}{3w-1},~\dot{\mathcal{T}}=-3H_{0}(1+w)\mathcal{T},~\ddot{\mathcal{T}}=9H_{0}^{2}(1+w)^{2}\mathcal{T}
\end{equation}
Using these values, the field equations (\ref{32}) and (\ref{33})
reduce to
\begin{equation}
\frac{\kappa^{2}\mathcal{T}}{3w-1}-\frac{1}{2}f+3H_{0}^{2}f_{R}+\frac{1+w}{3w-1}\mathcal{T}f_{\mathcal{T}}+12H_{0}^{4}f_{G}
+9H_{0}^{2}(1+w)\mathcal{T}f_{R\mathcal{T}}+36H_{0}^{4}(1+w)\mathcal{T}f_{G\mathcal{T}}=0
\end{equation}
and
$$
\frac{\kappa^{2}w\mathcal{T}}{3w-1}+\frac{1}{2}f-3H_{0}^{2}f_{R}-12H_{0}^{4}f_{G}
+3H_{0}^{2}(1+w)(1+3w)\mathcal{T}f_{R\mathcal{T}}+12H_{0}^{4}(1+w)(1+3w)\mathcal{T}f_{G\mathcal{T}}
$$
\begin{equation}
+9H_{0}^{2}(1+w)^{2}\mathcal{T}^{2}f_{R\mathcal{T}\mathcal{T}}+36H_{0}^{4}(1+w)^{2}\mathcal{T}^{2}f_{G\mathcal{T}\mathcal{T}}=0
\end{equation}
The solution of the above equations is obtained as
\begin{equation}
f(R,G,\mathcal{T})=c_{1}e^{a_{1}R+a_{2}G}\mathcal{T}^{b_{1}}+c_{2}e^{a_{3}R}\mathcal{T}^{b_{2}}+c_{3}e^{a_{4}G}\mathcal{T}^{b_{3}}
+c_{4}e^{a_{5}R}+c_{5}e^{a_{6}G}+c_{6}\mathcal{T}^{b_{4}}+c_{7}\mathcal{T}
\end{equation}
where $c_{i}~(i=1,...,7)$, $a_{i}~(i=1,...,6)$ and
$b_{i}~(i=1,...,4)$ are constants satisfying (i)
$a_{1}+4H_{0}^{2}a_{2}=-\frac{1}{42H_{0}^{2}},~a_{3}=-\frac{1}{42H_{0}^{2}},
~a_{4}=-\frac{1}{168H_{0}^{4}},a_{5}=\frac{1}{6H_{0}^{2}},~a_{6}=\frac{1}{24H_{0}^{4}},
~b_{1}=b_{2}=b_{3}=1,~c_{6}=0,~c_{7}=-\kappa^{2}$ for $w=1$; (ii)
$a_{1}+4H_{0}^{2}a_{2}=\frac{1}{6H_{0}^{2}},
~a_{3}=\frac{1}{6H_{0}^{2}},~a_{4}=\frac{1}{24H_{0}^{4}},a_{5}=\frac{1}{6H_{0}^{2}},~a_{6}=\frac{1}{24H_{0}^{4}},
~c_{6}=0,~c_{7}=-\frac{\kappa^{2}}{2}$ for $w=-1$.\\

So the above solution can be written as
\begin{equation}
f(R,G,\mathcal{T})=\left\{
\begin{array}{ll}
e^{-\frac{R}{42H_{0}^{2}}}\left(c_{1}e^{-4a_{2}H_{0}^{2}R+a_{2}G}+c_{2}\right)\mathcal{T}+c_{3}\mathcal{T}~e^{-\frac{G}{168H_{0}^{4}}}
+c_{4}e^{\frac{R}{6H_{0}^{2}}}+c_{5}e^{\frac{G}{24H_{0}^{4}}}-\kappa \mathcal{T}~~~for~~w=1\\
e^{\frac{R}{6H_{0}^{2}}}\left(c_{1}e^{-4a_{2}H_{0}^{2}R+a_{2}G}\mathcal{T}^{b_{1}}+c_{2}\mathcal{T}^{b_{2}}+c_{4}\right)+
e^{\frac{G}{24H_{0}^{4}}}\left(c_{3}\mathcal{T}^{b_{3}}+c_{5}\right)-\frac{\kappa^{2}\mathcal{T}}{2}~~~for~~w=-1
\end{array}
\right.
\end{equation}
We see that $f(R,G,\mathcal{T})$ is the combinations of
exponential and power forms of $R$, $G$ and $\mathcal{T}$. Now we
draw the function $f(R,G,\mathcal{T})$ against $t$ in figure 1.
From the figure we observe that $f(R,G,\mathcal{T})$ increases as
$t$ increases. From the figure we observe that
$f(R,G,\mathcal{T})$ sharply increases as $t$ increases (upto
$\approx 2$) and then it takes the value 5.4365 which is nearly
parallel to $t$ axis (i.e., slope of the curve $\approx 0$)
throughout the evolution.

\begin{figure}
\includegraphics[height=2.0in]{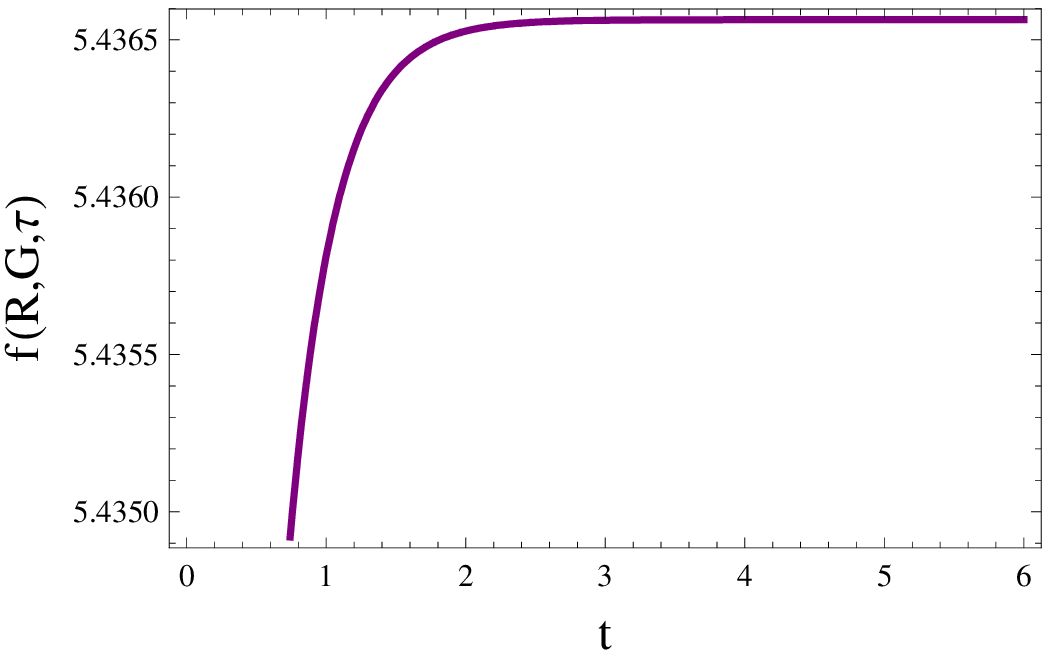}
\vspace{4mm}

Fig.1 : Plot of $f(R,G,\mathcal{T})$ against time $t$ for de
Sitter expansion. \vspace{0.2in}
\end{figure}

\section{Construction of $f(R,G,\mathcal{T})$ in Power Law Model}

Now we consider the universe expands in the power law form of the
scale factor \cite{Sharif}
\begin{equation}\label{51}
a(t)=a_{0}t^{n}
\end{equation}
where $a_{0}$ and $n$ are positive constants. It should be noted
that for acceleration phase of the Universe $\ddot{a}>0$, we must
have $n>1$. So from (\ref{51}) we obtain
\begin{equation}\label{52}
H=\frac{n}{t},~\dot{H}=-\frac{n}{t^{2}},~R=\frac{6n(2n-1)}{t^{2}},~\frac{\dot{R}}{R}=-\frac{2}{t},~
\frac{\ddot{R}}{R}=\frac{6}{t^{2}},
~G=\frac{24n^{3}(n-1)}{t^{4}},~\frac{\dot{G}}{G}=-\frac{4}{t},~\frac{\ddot{G}}{G}=\frac{20}{t^{2}}
\end{equation}
and
\begin{equation}\label{53}
\rho=\frac{\mathcal{T}}{3w-1},~\frac{\dot{\mathcal{T}}}{\mathcal{T}}=-\frac{3n(1+w)}{t},~\frac{\ddot{\mathcal{T}}}{\mathcal{T}}=\frac{3n(1+w)[3n(1+w)+1]}{t^{2}}
\end{equation}
Using equations (\ref{52}) and (\ref{53}), the field equation
(\ref{32}) simplifies to the form
\begin{eqnarray*}
\frac{\kappa^{2}\mathcal{T}}{3w-1}-\frac{1}{2}~f+\frac{n-1}{2(2n-1)}~Rf_{R}+\frac{1}{2}~Gf_{G}
+\frac{1+w}{3w-1}~\mathcal{T}f_{\mathcal{T}}+\frac{1}{2n-1}~R^{2}f_{RR}
\end{eqnarray*}
\begin{equation}\label{54}
+\frac{4n-3}{(n-1)(2n-1)}~RGf_{RG}+\frac{3n(1+w)}{2(2n-1)}~R\mathcal{T}f_{R\mathcal{T}}+\frac{2}{n-1}~G^{2}f_{GG}
+\frac{3n(1+w)}{2(n-1)}~G\mathcal{T}f_{G\mathcal{T}}=0
\end{equation}
From the above equation (\ref{54}), we get the solution  as in the
following form:
\begin{equation}
f(R,G,\mathcal{T})=c_{1}R^{a_{1}}G^{a_{2}}\mathcal{T}^{a_{3}}+c_{2}R^{a_{4}}G^{a_{5}}+c_{3}R^{a_{6}}\mathcal{T}^{a_{7}}
+c_{4}G^{a_{8}}\mathcal{T}^{a_{9}}+c_{5}R^{b_{1}}+c_{6}G^{b_{2}}+c_{7}\mathcal{T}^{b_{3}}+c_{8}\mathcal{T}+c_{9}G
\end{equation}
where $c_{i}~(i=1,...,9)$, $a_{i}~(i=1,...,9)$, $b_{i}~(i=1,2,3)$
are constants satisfying
\begin{equation}
c_{8}=\frac{2\kappa^{2}}{w-3},~~b_{3}=\frac{3w-1}{2(1+w)},~~b_{2}=-\frac{n-1}{4},~2b_{1}^{2}+(n-3)b_{1}=2n-1,
\end{equation}
\begin{equation}
a_{9}=\frac{(3w-1)(1-a_{8})(n-1+4a_{8})}{(1+w)[2(n-1)+3n(3w-1)a_{8}]}~,
~a_{7}=\frac{(3w-1)[2n-1-a_{6}(n-1-2a_{6})]}{(1+w)[2(2n-1)+3n(3w-1)a_{6}]}~,
\end{equation}
\begin{equation}
[4(2n-1)a_{5}+2(4n-3)a_{4}+(n-5)(2n-1)]a_{5}=(n-1)[2n-1-a_{4}(n-3+2a_{4})],
\end{equation}
\begin{equation}
a_{3}=\frac{(3w-1)[(n+1)\{(2n+1)(1-a_{2})-a_{1}(n+3-2a_{1})\}+2(4n+3)a_{1}a_{2}+4(2n+1)a_{2}(a_{2}-1)]}
{(1+w)[2(n+1)(2n+1)+3n(3w-1)\{(n+1)a_{1}+(2n+1)a_{2}\}]}
\end{equation}
We see that $f(R,G,\mathcal{T})$ is the power forms of $R$, $G$
and $\mathcal{T}$. Now we draw the function $f(R,G,\mathcal{T})$
against $t$ in figure 2. From the figure we observe that
$f(R,G,\mathcal{T})$ decreases as $t$ increases. From the figure
we observe that $f(R,G,\mathcal{T})$ sharply decreases as $t$
increases (upto $\approx 2$) and then it is nearly parallel to $t$
axis (i.e., slope of the curve $\approx 0$) throughout the
evolution.

\begin{figure}
\includegraphics[height=2.0in]{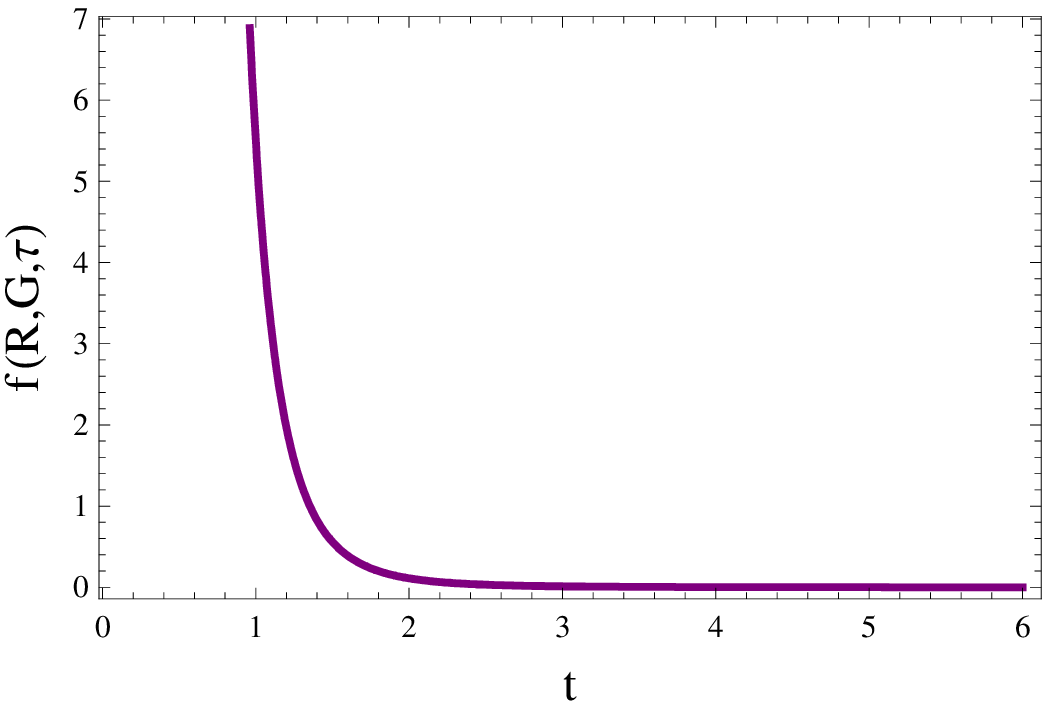}
\vspace{4mm}

Fig.2 : Plot of $f(R,G,\mathcal{T})$ against time $t$ for power
law expansion. \vspace{0.2in}
\end{figure}

\begin{figure}
\includegraphics[height=2.0in]{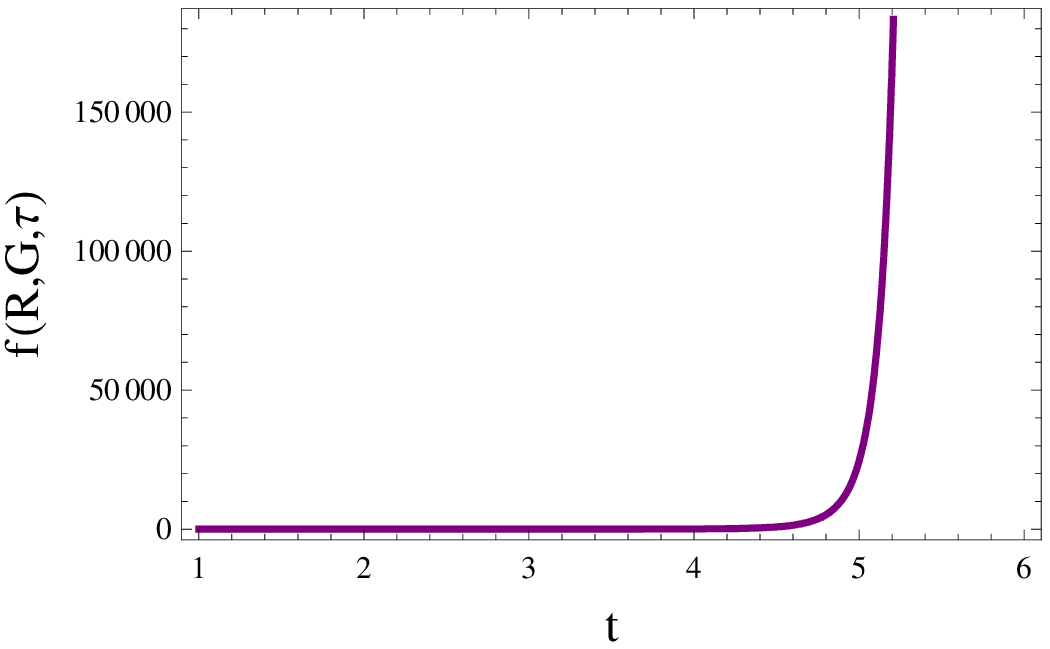}
\vspace{4mm}

Fig.3 : Plot of $f(R,G,\mathcal{T})$ against time $t$ for future
singularity model. \vspace{0.2in}
\end{figure}

\section{Construction of $f(R,G,\mathcal{T})$ in Future Singularity Model}

If there is a finite time future singularity occurs at $t=t_{s}$
in the evolution of the Universe, we may consider the universe
expands as in the following form of the scale factor \cite{Sharif}
\begin{equation}
a(t)=\frac{a_{0}}{(t_{s}-t)^{n}}
\end{equation}
where $a_{0}$ and $n$ are positive constants and $t<t_{s}$. Now in
this case we obtain $H=n/(t_{s}-t)$. Since $H$ becomes singular in
the limit $t\rightarrow t_{s}$ so $t_{s}$ is the future time when
a singularity appears. This singularity is known as type I or big
rip singularity \cite{N} because $a(t)\rightarrow \infty$,
$\rho\rightarrow \infty$ and $|p|\rightarrow \infty$ in the limit
$t\rightarrow t_{s}$. Similar to the power law form, the field
equation (\ref{32}) reduces to
\begin{eqnarray*}
\frac{\kappa^{2}\mathcal{T}}{3w-1}-\frac{1}{2}~f+\frac{n+1}{2(2n+1)}~Rf_{R}+\frac{1}{2}~Gf_{G}
+\frac{1+w}{3w-1}~\mathcal{T}f_{\mathcal{T}}-\frac{1}{2n+1}~R^{2}f_{RR}
\end{eqnarray*}
\begin{equation}
-\frac{4n+3}{(n+1)(2n+1)}~RGf_{RG}+\frac{3n(1+w)}{2(2n+1)}~R\mathcal{T}f_{R\mathcal{T}}-\frac{2}{n+1}~G^{2}f_{GG}
+\frac{3n(1+w)}{2(n+1)}~G\mathcal{T}f_{G\mathcal{T}}=0
\end{equation}
From the above equation, we get the solution in the following
form:
\begin{equation}
f(R,G,\mathcal{T})=d_{1}R^{x_{1}}G^{x_{2}}\mathcal{T}^{x_{3}}+d_{2}R^{x_{4}}G^{x_{5}}+d_{3}R^{x_{6}}\mathcal{T}^{x_{7}}
+d_{4}G^{x_{8}}\mathcal{T}^{x_{9}}+d_{5}R^{y_{1}}+d_{6}G^{y_{2}}+d_{7}\mathcal{T}^{y_{3}}+d_{8}\mathcal{T}+d_{9}G
\end{equation}
where $d_{i}~(i=1,...,9)$, $x_{i}~(i=1,...,9)$, $y_{i}~(i=1,2,3)$
are constants satisfying
\begin{equation}
d_{8}=\frac{2\kappa^{2}}{w-3},~~y_{3}=\frac{3w-1}{2(1+w)},~~y_{2}=\frac{n+1}{4},~2y_{1}^{2}-(n+3)y_{1}+(2n+1)=0,
\end{equation}
\begin{equation}
x_{9}=\frac{(3w-1)(1-x_{8})(n+1-4x_{8})}{(1+w)[2(n+1)+3n(3w-1)x_{8}]}~,
~x_{7}=\frac{(3w-1)[2n+1-x_{6}(n+1-2x_{6})]}{(1+w)[2(2n+1)+3n(3w-1)x_{6}]}~,
\end{equation}
\begin{equation}
[4(2n+1)x_{5}+2(4n+3)x_{4}-(n+5)(2n+1)]x_{5}=(n+1)[x_{4}(n+3-2x_{4})-2n+1],
\end{equation}
\begin{equation}
x_{3}=\frac{(3w-1)[(n-1)\{(2n-1)(1-x_{2})-x_{1}(n-3+2x_{1})\}-2(4n-3)x_{1}x_{2}-4(2n-1)x_{2}(x_{2}-1)]}
{(1+w)[2(n-1)(2n-1)+3n(3w-1)\{(n-1)x_{1}+(2n-1)x_{2}\}]}
\end{equation}
When the universe expands with big rip singularity, the solution
$f(R,G,\mathcal{T})$ is in the combinations of power forms of
$R,~G$ and $\mathcal{T}$. We see that $f(R,G,\mathcal{T})$ is the
power forms of $R$, $G$ and $\mathcal{T}$. Now we draw the
function $f(R,G,\mathcal{T})$ against $t$ in figure 3. From the
figure we observe that $f(R,G,\mathcal{T})$ nearly parallel to $t$
axis (i.e., slope of the curve $\approx 0$) upto certain period of
time $t\approx 5$ then sharply increases as $t$ increases near
future singularity ($t\approx 6$).

\begin{figure}
\includegraphics[height=2.0in]{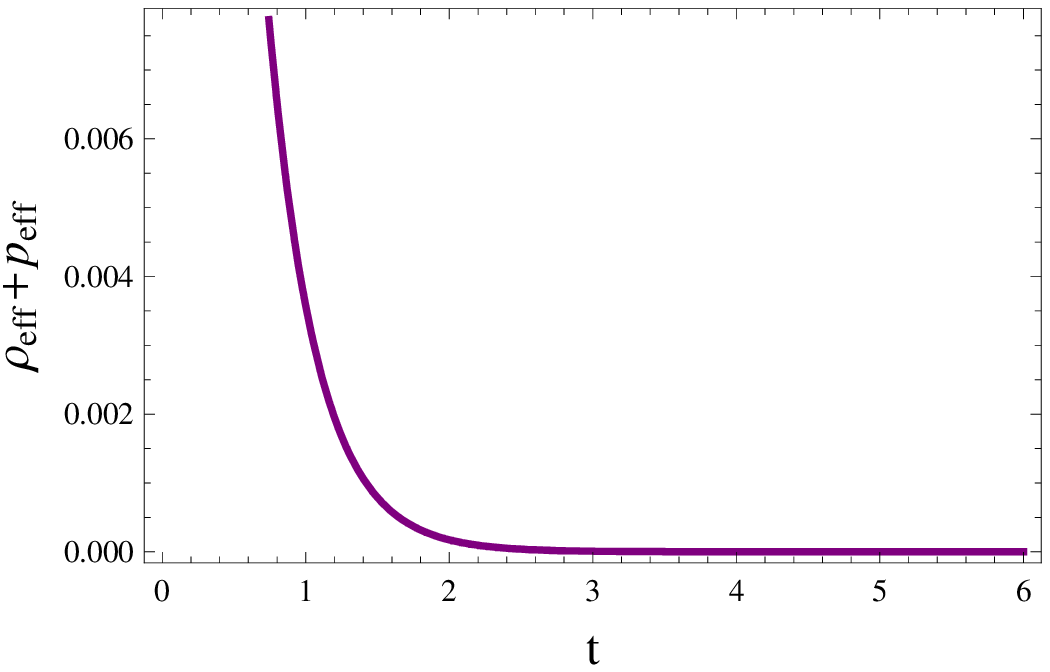}~~~~~
\includegraphics[height=2.0in]{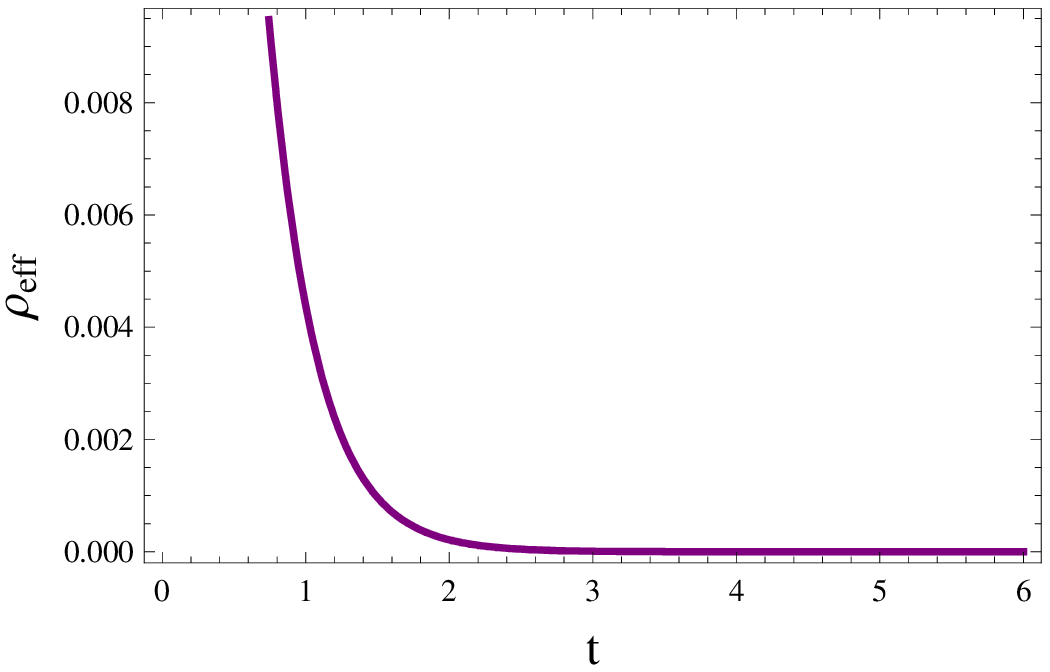}\\
\vspace{4mm}
~~~~~~~~~Fig.4~~~~~~~~~~~~~~~~~~~~~~~~~~~~~~~~~~~~~~~~~~~~~~~~~~~~~~~~~~~~~~~~~~~~~~~~~~~~~~~Fig.5\\
\vspace{4mm} Figs.4 and 5 show the plots of $\rho_{eff}+p_{eff}$
and $\rho_{eff}$
against time $t$ for de Sitter expansion model respectively.\\
\vspace{0.2in}

\includegraphics[height=2.0in]{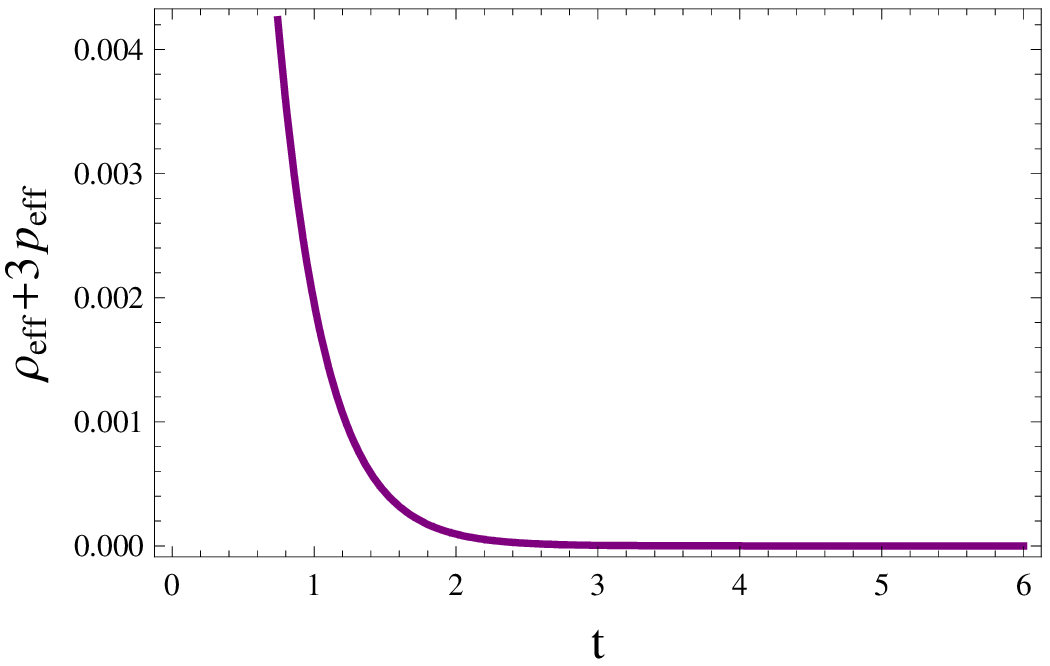}~~~~~
\includegraphics[height=2.0in]{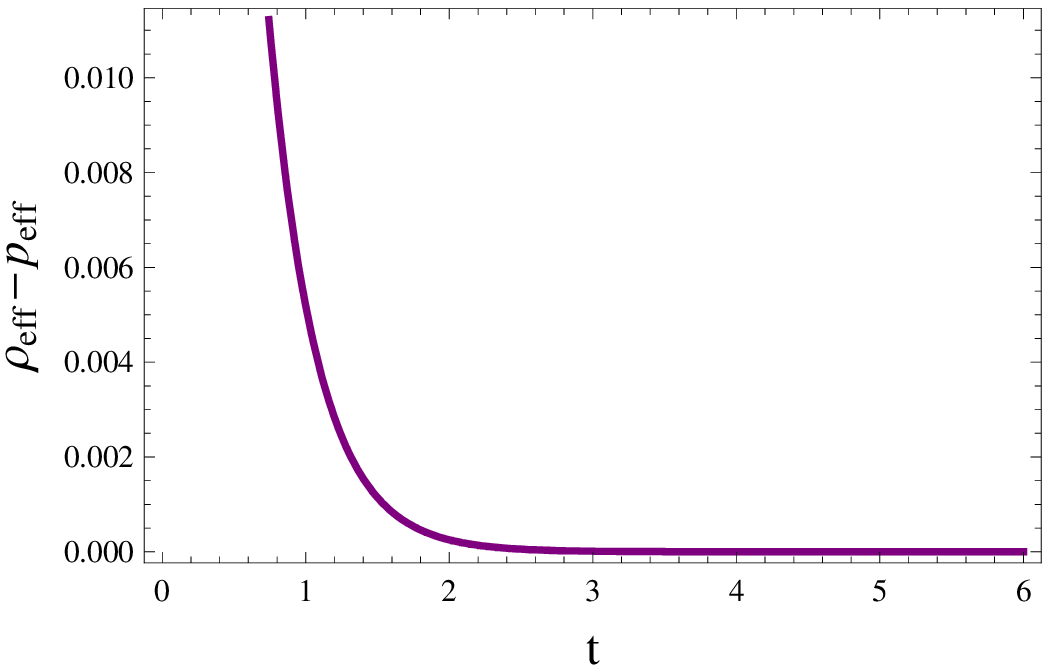}\\
\vspace{4mm}
~~~~~~~~~Fig.6~~~~~~~~~~~~~~~~~~~~~~~~~~~~~~~~~~~~~~~~~~~~~~~~~~~~~~~~~~~~~~~~~~~~~~~~~~~~~~~Fig.7\\
\vspace{4mm} Figs.6 and 7 show the plots of $\rho_{eff}+3p_{eff}$
and $\rho_{eff}-p_{eff}$ against time $t$ for de Sitter expansion
model respectively. \vspace{0.2in}
\end{figure}

\begin{figure}
\includegraphics[height=2.0in]{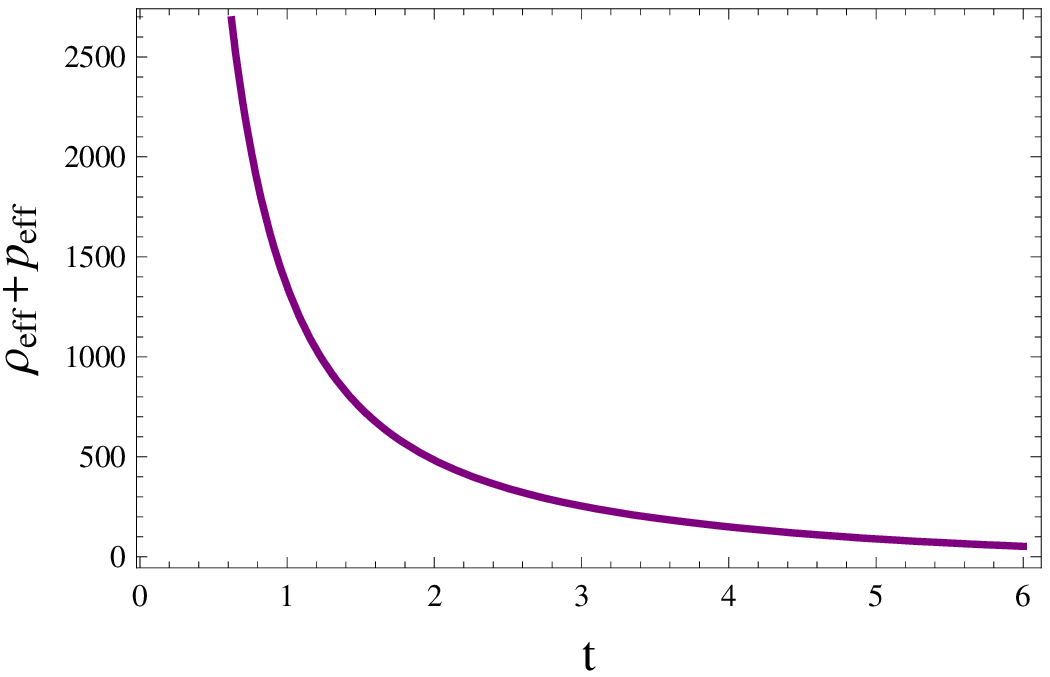}~~~~~
\includegraphics[height=2.0in]{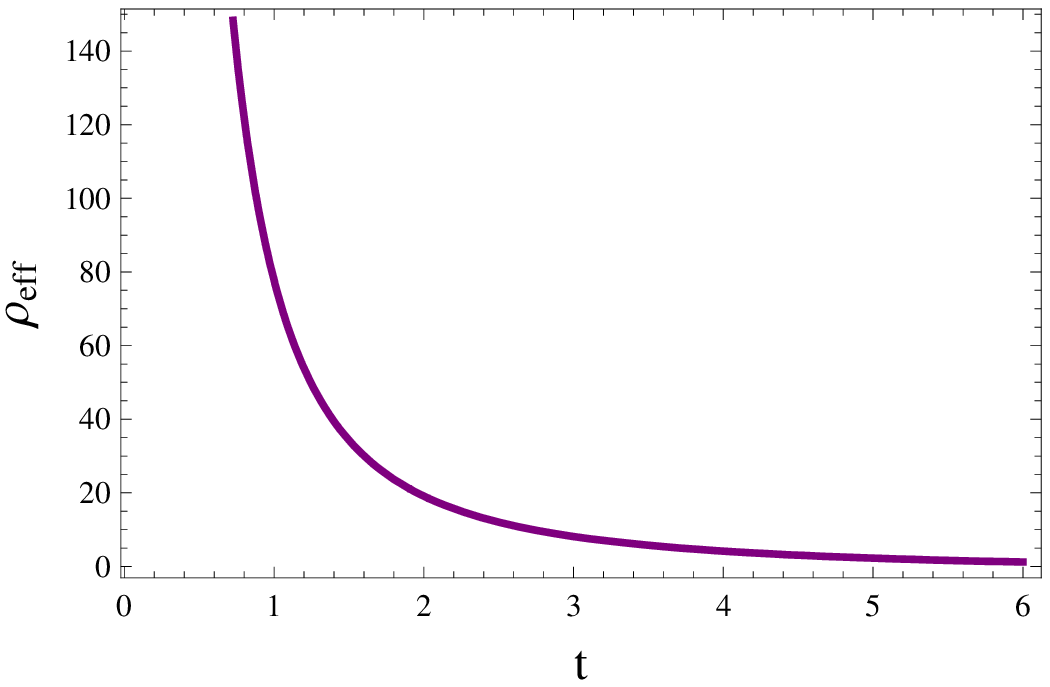}\\
\vspace{4mm}
~~~~~~~~~Fig.8~~~~~~~~~~~~~~~~~~~~~~~~~~~~~~~~~~~~~~~~~~~~~~~~~~~~~~~~~~~~~~~~~~~~~~~~~~~~~~~Fig.9\\
\vspace{4mm} Figs.8 and 9 show the plots of $\rho_{eff}+p_{eff}$
and $\rho_{eff}$
against time $t$ for power law expansion model respectively.\\
\vspace{0.2in}

\includegraphics[height=2.0in]{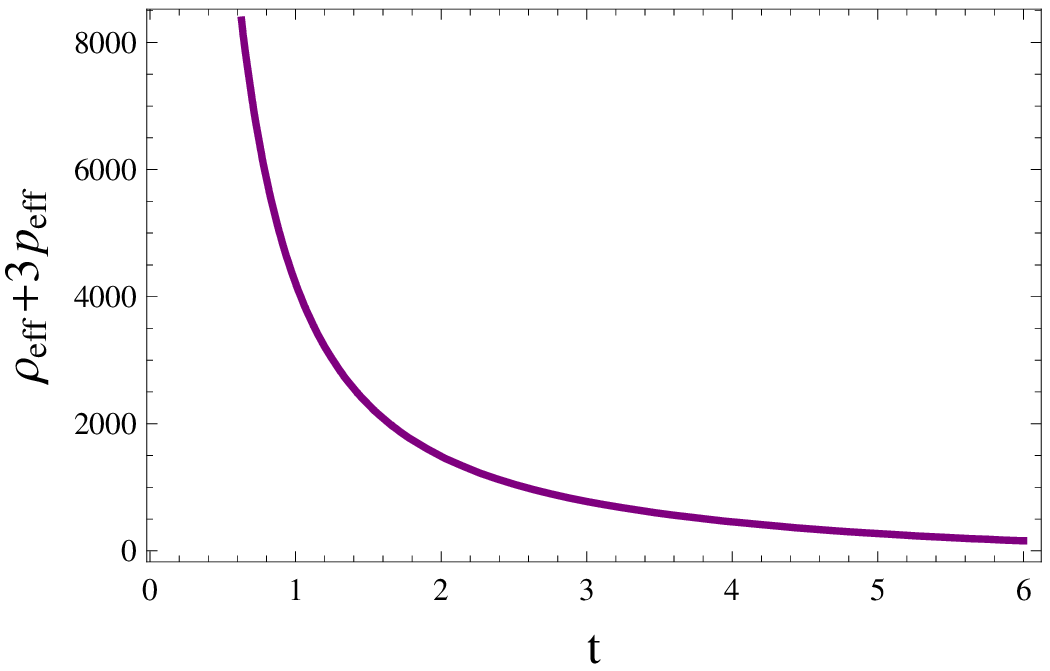}~~~~~
\includegraphics[height=2.0in]{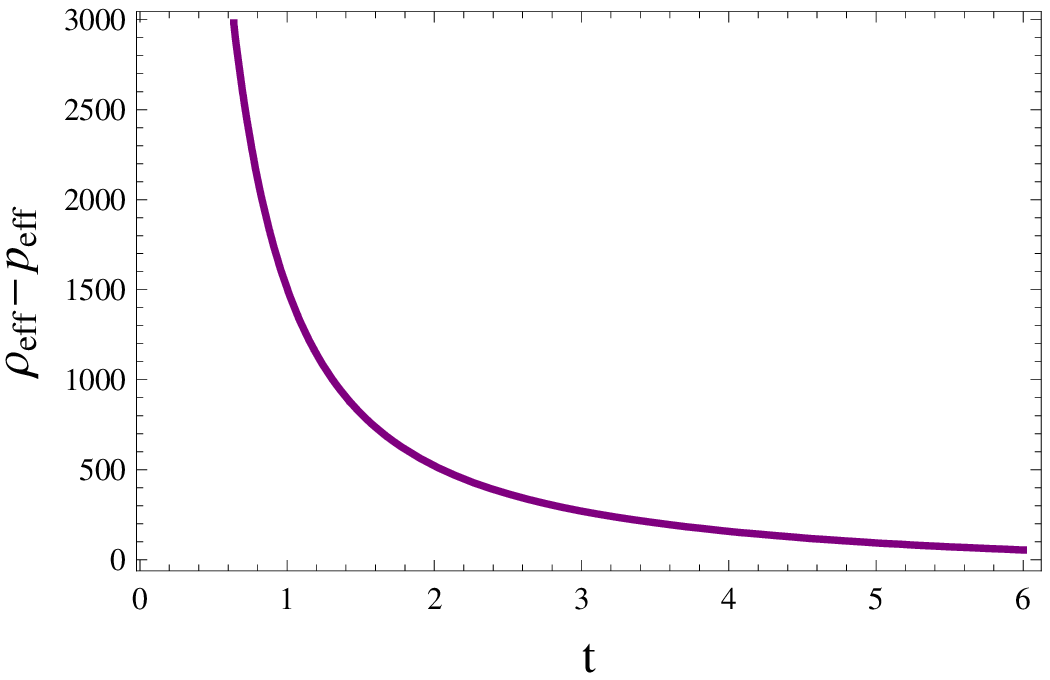}\\
\vspace{4mm}
~~~~~~~~~Fig.10~~~~~~~~~~~~~~~~~~~~~~~~~~~~~~~~~~~~~~~~~~~~~~~~~~~~~~~~~~~~~~~~~~~~~~~~~~~~~~~Fig.11\\
\vspace{4mm} Figs.10 and 11 show the plots of
$\rho_{eff}+3p_{eff}$ and $\rho_{eff}-p_{eff}$ against time $t$
for power law expansion model respectively. \vspace{0.2in}
\end{figure}

\begin{figure}
\includegraphics[height=2.0in]{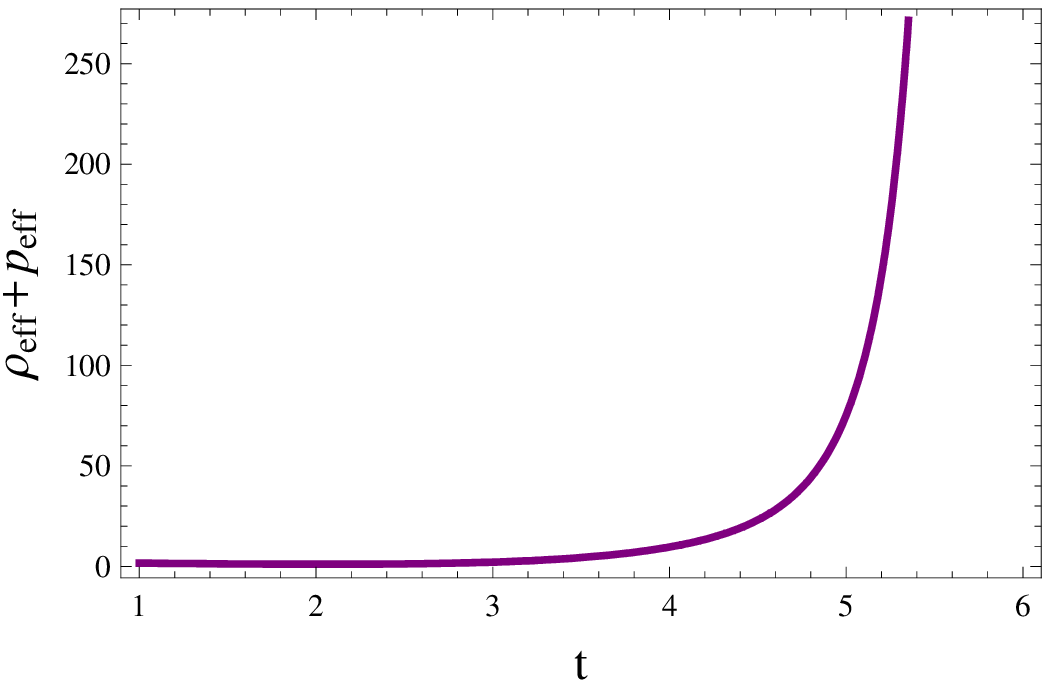}~~~~~
\includegraphics[height=2.0in]{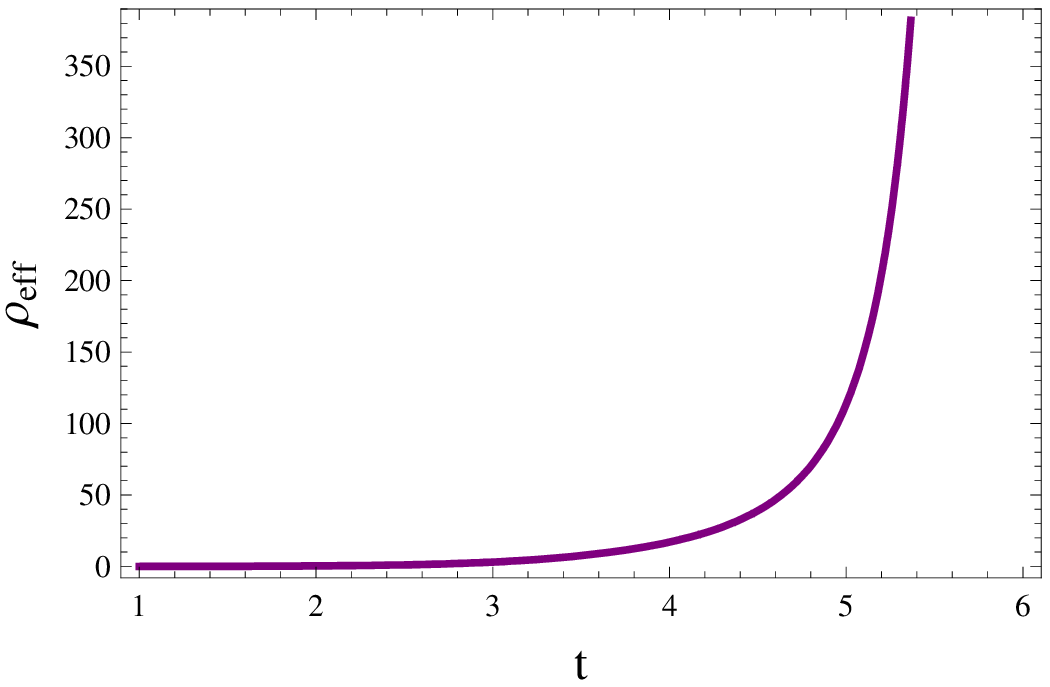}\\
\vspace{4mm}
~~~~~~~~~Fig.12~~~~~~~~~~~~~~~~~~~~~~~~~~~~~~~~~~~~~~~~~~~~~~~~~~~~~~~~~~~~~~~~~~~~~~~~~~~~~~~Fig.13\\
\vspace{4mm} Figs.12 and 13 show the plots of $\rho_{eff}+p_{eff}$
and $\rho_{eff}$
against time $t$ for future singularity model respectively.\\
\vspace{0.2in}

\includegraphics[height=2.0in]{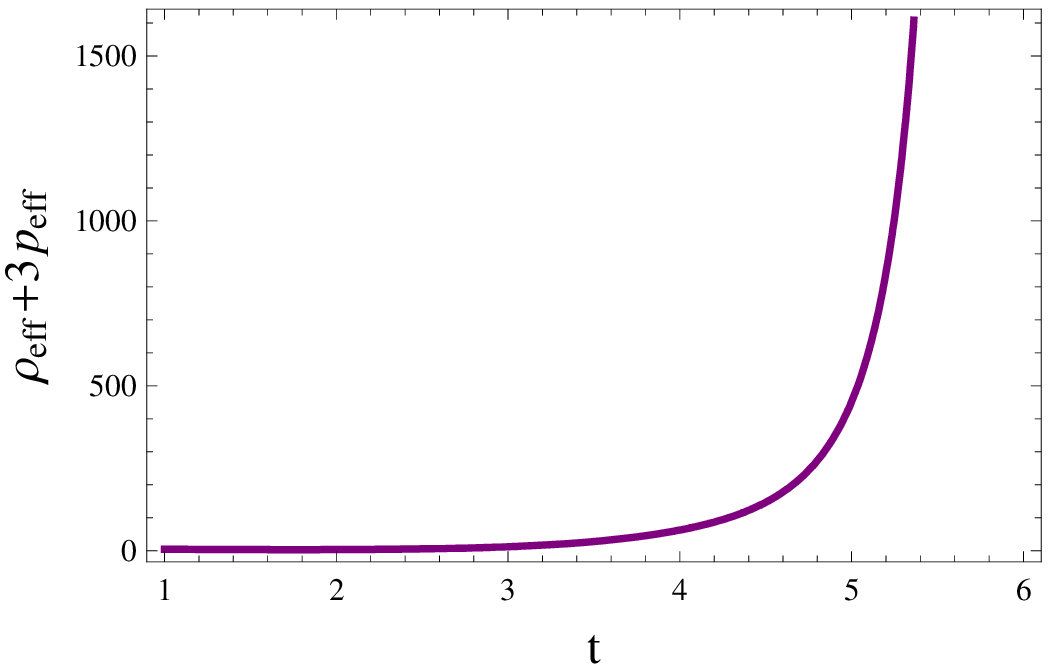}~~~~~
\includegraphics[height=2.0in]{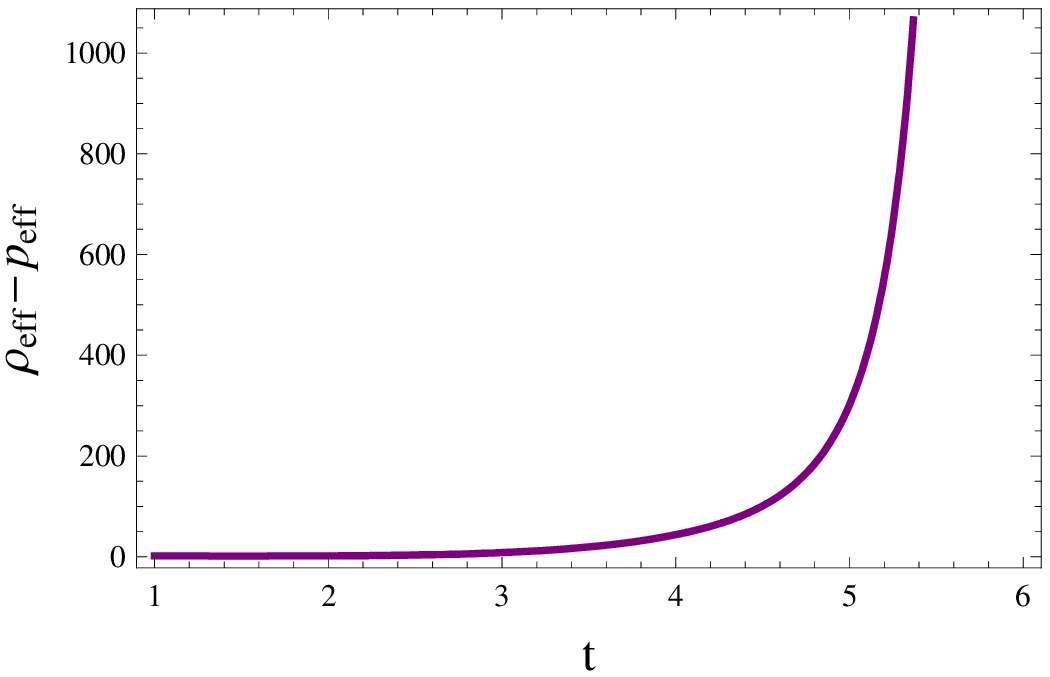}\\
\vspace{4mm}
~~~~~~~~~Fig.14~~~~~~~~~~~~~~~~~~~~~~~~~~~~~~~~~~~~~~~~~~~~~~~~~~~~~~~~~~~~~~~~~~~~~~~~~~~~~~~Fig.15\\
\vspace{4mm} Figs.14 and 15 show the plots of
$\rho_{eff}+3p_{eff}$ and $\rho_{eff}-p_{eff}$ against time $t$
for future singularity model respectively. \vspace{0.2in}
\end{figure}

\section{Energy Conditions}

Here we study the energy conditions for $f(R,G,\mathcal{T})$
gravity in FRW Universe. The concept of energy conditions came
from the Raychaudhuri equation \cite{Ata,Sharif}
\begin{equation}
\frac{d\theta}{d\tau}=-\frac{1}{2}\theta^{2}-\sigma_{\mu\nu}\sigma^{\mu\nu}+\omega_{\mu\nu}\omega^{\mu\nu}
-R_{\mu\nu}k^{\mu}k^{\nu}
\end{equation}
where $\theta$, $\sigma_{\mu\nu}$ and $\omega_{\mu\nu}$ are
expansion scalar, shear tensor and rotation tensor respectively
associated to congruence defined by the null vector field
$k^{\mu}$. From the Raychaudhuri equation, we observe that
$\sigma^{2}=\sigma_{\mu\nu}\sigma^{\mu\nu} \ge 0$ and for any
hypersurface orthogonal congruences, $\omega_{\mu\nu}=0$ and for
attractive gravity, $\frac{d\theta}{d\tau} \le 0$, so above
equation reduces to $R_{\mu\nu}k^{\mu}k^{\nu}\ge 0$. In Einstein's
gravity, the above condition becomes $T_{\mu\nu}k^{\mu}k^{\nu} \ge
0$, which is the null energy condition (NEC). For timelike vector
field $v^{\mu}$, the above Raychaudhuri equation can be written as
\begin{equation}
\frac{d\theta}{d\tau}=-\frac{1}{3}\theta^{2}-\sigma_{\mu\nu}\sigma^{\mu\nu}+\omega_{\mu\nu}\omega^{\mu\nu}
-R_{\mu\nu}v^{\mu}v^{\nu}
\end{equation}
From this equation we get $R_{\mu\nu}v^{\mu}v^{\nu} \ge 0$ and
hence $T_{\mu\nu}v^{\mu}v^{\nu} \ge 0$, which is the weak energy
condition (WEC). Similarly the strong energy condition (SEC) is
$(T_{\mu\nu}-\frac{1}{2}\mathcal{T}g_{\mu\nu})v^{\mu}v^{\nu} \ge
0$ and the dominant energy condition (DEC) is
$T_{\mu\nu}v^{\mu}v^{\nu} \ge 0$ and $T_{\mu\nu}v^{\nu}$ is not
space-like imply locally measured energy density to be always
positive and the energy flux is time-like or null.
Now we write the NEC, WEC, SEC and DEC for modified $f(R,G,\mathcal{T})$ gravity theory in general values of $f(R,G,\mathcal{T})$.\\

$\bullet$ Null energy condition (NEC):
\begin{equation}
\rho_{eff}+p_{eff}=\frac{1}{\kappa^{2}f_{R}} \left[
\kappa^{2}(\rho+p)+(\rho+p)f_{\mathcal{T}}-H\dot{f}_{R}+\ddot{f}_{R}
+4H(2\dot{H}-H^{2})\dot{f}_{G}+4H^{2}\ddot{f}_{G} \right] \ge 0,
\end{equation}

$\bullet$ Weak energy condition (WEC):
\begin{equation}
\rho_{eff}=\frac{1}{\kappa^{2}f_{R}} \left[
\kappa^{2}\rho+(\rho+p)f_{\mathcal{T}}+\frac{1}{2}(Rf_{R}-f)-3H\dot{f}_{R}+12H^{2}(\dot{H}+H^{2})f_{G}-12H^{3}\dot{f}_{G}
\right] \ge 0,
\end{equation}
\begin{equation}
\rho_{eff}+p_{eff}=\frac{1}{\kappa^{2}f_{R}} \left[
\kappa^{2}(\rho+p)+(\rho+p)f_{\mathcal{T}}-H\dot{f}_{R}+\ddot{f}_{R}
+4H(2\dot{H}-H^{2})\dot{f}_{G}+4H^{2}\ddot{f}_{G} \right] \ge 0,
\end{equation}

$\bullet$ Strong energy condition (SEC):
\begin{eqnarray*}
\rho_{eff}+3p_{eff}=\frac{1}{\kappa^{2}f_{R}} \left[
\kappa^{2}(\rho+3p)+(\rho+p)f_{\mathcal{T}}-(Rf_{R}-f)+3H\dot{f}_{R}+3\ddot{f}_{R}\right.
\end{eqnarray*}
\begin{equation}
~~~~~~~~~~~~~~~~~~~~~~\left. -24H^{2}(\dot{H}+H^{2})f_{G}
+12H(2\dot{H}+H^{2})\dot{f}_{G}+12H^{2}\ddot{f}_{G} \right] \ge 0,
\end{equation}
\begin{equation}
\rho_{eff}+p_{eff}=\frac{1}{\kappa^{2}f_{R}} \left[
\kappa^{2}(\rho+p)+(\rho+p)f_{\mathcal{T}}-H\dot{f}_{R}+\ddot{f}_{R}
+4H(2\dot{H}-H^{2})\dot{f}_{G}+4H^{2}\ddot{f}_{G} \right] \ge 0
\end{equation}

$\bullet$ Dominant energy condition (DEC):
\begin{equation}
\rho_{eff}=\frac{1}{\kappa^{2}f_{R}} \left[
\kappa^{2}\rho+(\rho+p)f_{\mathcal{T}}+\frac{1}{2}(Rf_{R}-f)-3H\dot{f}_{R}+12H^{2}(\dot{H}+H^{2})f_{G}-12H^{3}\dot{f}_{G}
\right] \ge 0,
\end{equation}
\begin{equation}
\rho_{eff}+p_{eff}=\frac{1}{\kappa^{2}f_{R}} \left[
\kappa^{2}(\rho+p)+(\rho+p)f_{\mathcal{T}}-H\dot{f}_{R}+\ddot{f}_{R}
+4H(2\dot{H}-H^{2})\dot{f}_{G}+4H^{2}\ddot{f}_{G} \right] \ge 0,
\end{equation}
\begin{eqnarray*}
\rho_{eff}-p_{eff}=\frac{1}{\kappa^{2}f_{R}} \left[
\kappa^{2}(\rho-p)+(\rho+p)f_{\mathcal{T}}+(Rf_{R}-f)-5H\dot{f}_{R}-\ddot{f}_{R}\right.
\end{eqnarray*}
\begin{equation}
~~~~~~~~~~~~~~~~~~~~~~~\left. +24H^{2}(\dot{H}+H^{2})f_{G}
-4H(2\dot{H}+5H^{2})\dot{f}_{G}-4H^{2}\ddot{f}_{G} \right] \ge 0
\end{equation}

Since the above expressions for NEC, WEC, SEC and DEC are
complicated form which contain partial derivatives of
$f(R,G,\mathcal{T})$ w.r.t. $R$, $G$ and $\mathcal{T}$ as well as
time derivatives. So to examine the validities of NEC, WEC, SEC
and DEC, we need graphical representations. For de Sitter model,
we plot $\rho_{eff}+p_{eff}$, $\rho_{eff}$, $\rho_{eff}+3p_{eff}$
and $\rho_{eff}-p_{eff}$ against $t$ in figures 4, 5, 6 and 7
respectively. We have taken the parameters
$H_{0}=72,~a_{0}=1,~\rho_{0}=1,~\kappa=1,a_{2}=1,~b_{1}=2,~b_{2}=3,~b_{3}=1,
b_{4}=2,c_{1}=2,~c_{2}=1,c_{3}=4,~c_{5}=3,~c_{6}=0$. For these
choices of the parameters, from figures we observe that
$\rho_{eff}+p_{eff}$, $\rho_{eff}$, $\rho_{eff}+3p_{eff}$ and
$\rho_{eff}-p_{eff}$ are all positive during evolution of the
Universe. So NEC, WEC, SEC and DEC are satisfied for our
constructed $f(R,G,\mathcal{T})$ gravity for se Sitter expansion
model. Also, we see that in this model, $\rho_{eff}+p_{eff}$,
$\rho_{eff}$, $\rho_{eff}+3p_{eff}$ and $\rho_{eff}-p_{eff}$ are
all decreasing as time increases. Initially, these quantities have
sharp decrease bahaviour (about $t\approx 2$) and after that these
are nearly parallel to $t$ axis and tending to zero but keeps
positive sign.

For power law model, we plot $\rho_{eff}+p_{eff}$, $\rho_{eff}$,
$\rho_{eff}+3p_{eff}$ and $\rho_{eff}-p_{eff}$ against $t$ in
figures 8, 9, 10 and 11 respectively. We have taken the parameters
$n=3,~a_{0}=1,~\rho_{0}=1,
~\kappa=1,a_{1}=2,~a_{2}=1,a_{4}=3,~a_{5}=2,~a_{6}=3,a_{8}=2,~b_{1}=2,~c_{1}=1,~c_{2}=3,c_{3}=2,~c_{5}=4,~c_{5}=1,~
c_{6}=1,~c_{7}=3,c_{9}=2$. For these choices of the parameters,
from figures we observe that $\rho_{eff}+p_{eff}$, $\rho_{eff}$,
$\rho_{eff}+3p_{eff}$ and $\rho_{eff}-p_{eff}$ are all positive
during evolution of the Universe. So NEC, WEC, SEC and DEC are
satisfied for our constructed $f(R,G,\mathcal{T})$ gravity for
power law expansion model. Also, we see that in this model,
$\rho_{eff}+p_{eff}$, $\rho_{eff}$, $\rho_{eff}+3p_{eff}$ and
$\rho_{eff}-p_{eff}$ are all decreasing from high values to nearly
zero as time increases.

For future singularity model, we plot $\rho_{eff}+p_{eff}$,
$\rho_{eff}$, $\rho_{eff}+3p_{eff}$ and $\rho_{eff}-p_{eff}$
against $t$ in figures 12, 13, 14 and 15 respectively. We have
taken the parameters $n=2,~a_{0}=1,~\rho_{0}=1,
~\kappa=1,~t_{s}=6,~x_{1}=2,~x_{2}=1,x_{4}=3,~x_{5}=2,~x_{6}=3,x_{8}=2,~y_{1}=2,~d_{1}=1,~d_{2}=3,d_{3}=2,~d_{5}=4,~d_{5}=1,~
d_{6}=1,~d_{7}=3,d_{9}=2$. For these choices of the parameters,
from figures we observe that $\rho_{eff}+p_{eff}$, $\rho_{eff}$,
$\rho_{eff}+3p_{eff}$ and $\rho_{eff}-p_{eff}$ are all positive
during evolution of the Universe. So NEC, WEC, SEC and DEC are
satisfied for our constructed $f(R,G,\mathcal{T})$ gravity for
future singularity model. Also, we see that in this model,
$\rho_{eff}+p_{eff}$, $\rho_{eff}$, $\rho_{eff}+3p_{eff}$ and
$\rho_{eff}-p_{eff}$ are all increasing from lower values to
higher values as time increases. At future singularity, all the
quantities are blows up.

\section{Stability Analysis}

In this section, we test the viability of $f(R,G,\mathcal{T})$
gravity model by exploring its stability against perturbation. The
square speed of sound is the key quantity for investigation of the
stability. So for stability analysis of the model against small
perturbation we derive the squared speed of sound defined by
\begin{equation}
v_{s}^{2}=\frac{dp_{eff}}{d\rho_{eff}}=\frac{\dot{p}_{eff}}{\dot{\rho}_{eff}}
\end{equation}
The sign of $v_{s}^{2}$ plays a crucial role in determining the
classical stability or instability of the background evolution. If
$0<v_{s}^{2}<1$, the model is classically stable while
$v_{s}^{2}<0$ or $v_{s}^{2}>1$ represent a classically unstable
model against the perturbation respectively. The square speed of
light $v_{s}^{2}$ vs time $t$ for de Sitter, power law and future
singularity models have been drawn in figures 16, 17 and 18
respectively. From these figures, we observe that $v_{s}^{2}$ for
all the models are lying in $(0,1)$. So we may conclude that
$f(R,G,\mathcal{T})$ gravity models are classically stable for de
Sitter, power law and future singularity expansions.

\begin{figure}
\includegraphics[height=2.0in]{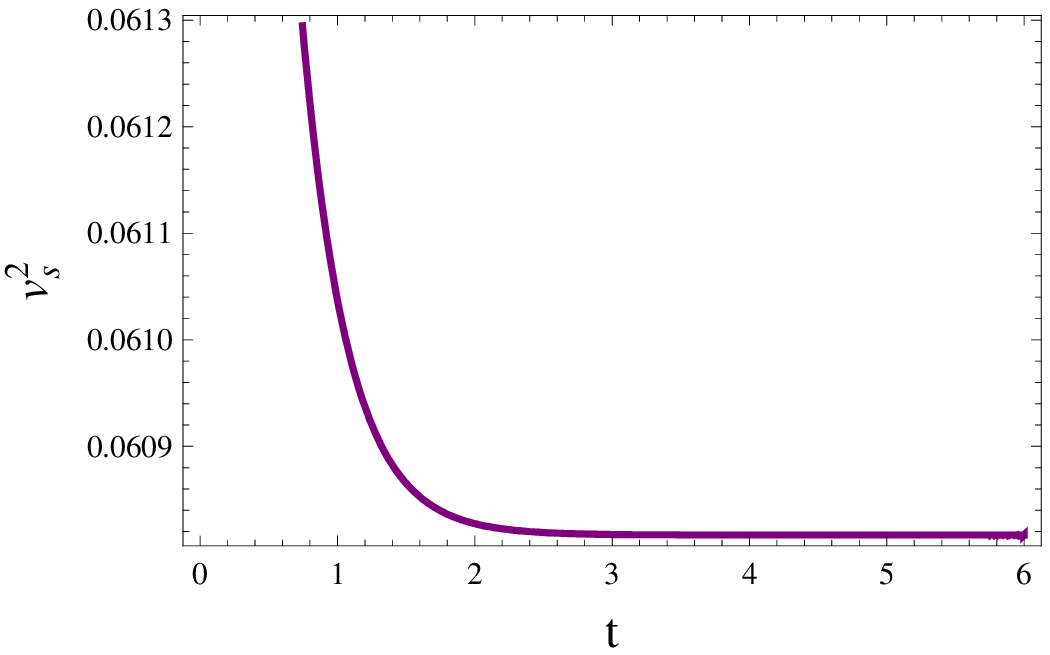}
\vspace{4mm}

Fig.16 : Plot of $v_{s}^{2}$ against time $t$ for de Sitter
expansion. \vspace{0.2in}
\end{figure}

\begin{figure}
\includegraphics[height=2.0in]{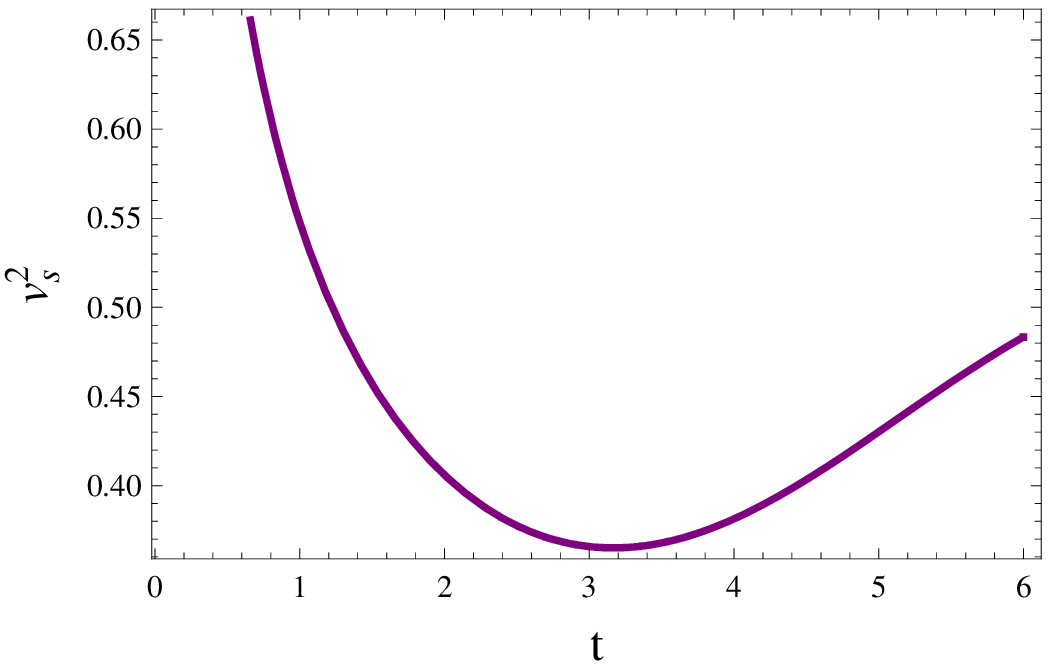}
\vspace{4mm}

Fig.17 : Plot of $v_{s}^{2}$ against time $t$ for power law
expansion. \vspace{0.2in}
\end{figure}

\begin{figure}
\includegraphics[height=2.0in]{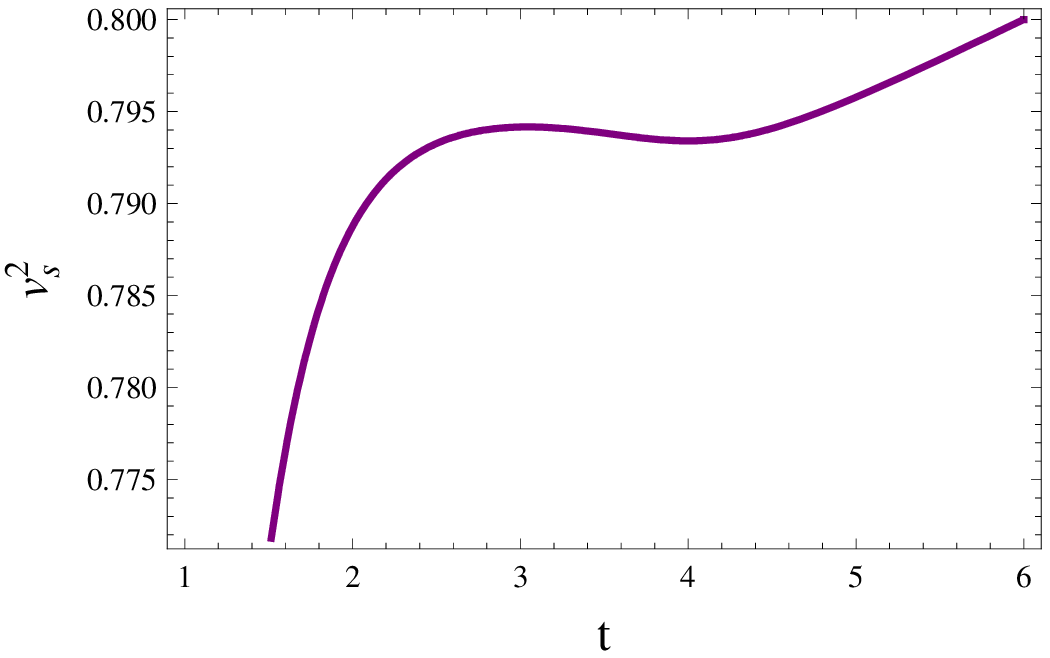}
\vspace{4mm}

Fig.18 : Plot of $v_{s}^{2}$ against time $t$ for future
singularity model. \vspace{0.2in}
\end{figure}

\section{Discussions and Concluding Remarks}

Here we have introduced the extended modified gravity theory named
as $f(R,G,\mathcal{T})$ gravity after the modifications of
gravities like $f(R,\mathcal{T}),~F(R,G),~f(G,\mathcal{T})$
gravities where $R$ is the Ricci scalar, $G$ is the Gauss-Bonnet
invariant and $\mathcal{T}$ is the trace of the stress-energy
tensor. We have obtained the gravitational lagrangian by adding
$f(R,G,\mathcal{T})$ with matter Lagrangian $L_{m}$ in the
Einstein-Hilbert action. We have derived the gravitational field
equations for $f(R,G,\mathcal{T})$ gravity by least action
principle. If we put $f(R,G,\mathcal{T})=f(R,\mathcal{T})$ ($G$
independent), we can recover the field equations in
$f(R,\mathcal{T})$ gravity which was proposed in Ref \cite{Harko}.
If we put $f(R,G,\mathcal{T}) =f(G,\mathcal{T})$ ($R$
independent), we can recover the field equations in
$f(G,\mathcal{T})$ gravity which was proposed in Ref \cite{Sharif}
and if we put  $f(R,G,\mathcal{T})=f(R,G)$ ($\mathcal{T}$
independent), we can recover the field equations in $f(R,G)$
gravity \cite{Bamba}. Next we have constructed the
$f(R,G,\mathcal{T})$ in terms of $R$, $G$ and $\mathcal{T}$ in de
Sitter as well as power law expansion of the universe. We have
also constructed $f(R,G,\mathcal{T})$ where the expansion follows
the finite time future singulary (big rip singularity). It has
been observed that for de Sitter expansion, the form of
$f(R,G,\mathcal{T})$ contains the combinations of exponential and
power forms of $R$, $G$ and $\mathcal{T}$ but for power law and
big rip singularity expansion models, the forms of
$f(R,G,\mathcal{T})$ contain only the power forms of $R$, $G$ and
$\mathcal{T}$.

We have drawn the function $f(R,G,\mathcal{T})$ against $t$ in
figure 1, 2 and 3 for de Sitter, power law and future singularity
models respectively. From figure 1, we have observed that
$f(R,G,\mathcal{T})$ sharply increases as $t$ increases (upto
$\approx 2$) and then it takes the value 5.4365 which is nearly
parallel to $t$ axis (i.e., slope of the curve $\approx 0$)
throughout the evolution of the Universe for de Sitter expansion.
From figure 2, we have seen that $f(R,G,\mathcal{T})$ sharply
decreases as $t$ increases (upto $\approx 2$) and then it is
nearly parallel to $t$ axis (i.e., slope of the curve $\approx 0$)
throughout the evolution of the Universe for power law expansion.
On the other hand, from figure 3, we have observed that
$f(R,G,\mathcal{T})$ nearly parallel to $t$ axis (i.e., slope of
the curve $\approx 0$) upto certain period of time $t\approx 5$
then sharply increases as $t$ increases near future singularity
($t\approx 6$).

We have investigated all the energy conditions (NEC, WEC, SEC,
DEC) in $f(R,G,\mathcal{T})$ modified theory of gravity. If
$\rho_{eff}+p_{eff}$, $\rho_{eff}$, $\rho_{eff}+3p_{eff}$ and
$\rho_{eff}-p_{eff}$ are all non-negative, then all the energy
conditions are satisfied. For this purpose, in de Sitter model, we
have plotted  $\rho_{eff}+p_{eff}$, $\rho_{eff}$,
$\rho_{eff}+3p_{eff}$ and $\rho_{eff}-p_{eff}$ against $t$ in
figures 4, 5, 6 and 7 respectively. From these figures we have
seen that $\rho_{eff}+p_{eff}$, $\rho_{eff}$,
$\rho_{eff}+3p_{eff}$ and $\rho_{eff}-p_{eff}$ are all positive
during evolution of the Universe. So NEC, WEC, SEC and DEC are
satisfied for our constructed $f(R,G,\mathcal{T})$ gravity for se
Sitter expansion model. Also, we have seen that in this model,
$\rho_{eff}+p_{eff}$, $\rho_{eff}$, $\rho_{eff}+3p_{eff}$ and
$\rho_{eff}-p_{eff}$ are all decreasing as time increases.
Initially, these quantities have sharp decrease bahaviour (about
$t\approx 2$) and after that these are nearly parallel to $t$ axis
and tending to zero but keeps positive sign. For power law model,
we have plotted $\rho_{eff}+p_{eff}$, $\rho_{eff}$,
$\rho_{eff}+3p_{eff}$ and $\rho_{eff}-p_{eff}$ against $t$ in
figures 8, 9, 10 and 11 respectively. From these figures, we have
observed that $\rho_{eff}+p_{eff}$, $\rho_{eff}$,
$\rho_{eff}+3p_{eff}$ and $\rho_{eff}-p_{eff}$ are all positive
during evolution of the Universe. So NEC, WEC, SEC and DEC are
satisfied for our constructed $f(R,G,\mathcal{T})$ gravity for
power law expansion model. Also, we have seen that in this model,
$\rho_{eff}+p_{eff}$, $\rho_{eff}$, $\rho_{eff}+3p_{eff}$ and
$\rho_{eff}-p_{eff}$ are all decreasing from high values to nearly
zero as time increases. Also for future singularity model, we have
plotted $\rho_{eff}+p_{eff}$, $\rho_{eff}$, $\rho_{eff}+3p_{eff}$
and $\rho_{eff}-p_{eff}$ against $t$ in figures 12, 13, 14 and 15
respectively. From figures we have seen  that
$\rho_{eff}+p_{eff}$, $\rho_{eff}$, $\rho_{eff}+3p_{eff}$ and
$\rho_{eff}-p_{eff}$ are all positive during evolution of the
Universe. So NEC, WEC, SEC and DEC are satisfied for our
constructed $f(R,G,\mathcal{T})$ gravity for future singularity
model. Also, we have seen that in this model,
$\rho_{eff}+p_{eff}$, $\rho_{eff}$, $\rho_{eff}+3p_{eff}$ and
$\rho_{eff}-p_{eff}$ are all increasing from lower values to
higher values as time increases. At future singularity, all the
quantities are blows up.

Finally, we have examined the stability of our constructed
$f(R,G,\mathcal{T})$ gravity for de Sitter, power law and future
singularity models. For this purpose, the sign of $v_{s}^{2}$
plays a crucial role to determine the classical stability or
instability of the background evolution. The square speed of light
$v_{s}^{2}$ vs time $t$ for de Sitter, power law and future
singularity models have been drawn in figures 16, 17 and 18
respectively. From these figures, we have seen that $v_{s}^{2}$
for all the models are lying in $(0,1)$. So we have concluded that
$f(R,G,\mathcal{T})$ gravity models are classically stable for
de Sitter, power law and future singularity expansions.\\\\

{\bf Acknowledgement}: The author is thankful to SERB DST (MATRICS
Scheme), Govt. of India for providing
research project grant (No. MTR/2019/000751/MS). \\\\

\end{document}